\documentclass[a4paper,11pt]{amsart}
\begin{document}

\title{On Einsteinian orbits of celestial bodies}
\author{Angelo Loinger}
\date{}
\address{Dipartimento di Fisica, Universit\`a di Milano, Via
Celoria 16 - 20133 Milano (Italy)}
\email{angelo.loinger@mi.infn.it}
\thanks{to be published on Spacetime \& Substance}

\maketitle

\begin{abstract} It can be demonstrated
that no motion of masses can generate gravitational waves.
Accordingly: \emph{i}) the time decrease of the orbital period of
the famous binary PSR1913+16 cannot yield an experimental proof of
the emission of gravitational waves; \emph{ii}) measurements of
the propagation of the quasar J0842+1835 radio-signals past
Jupiter cannot reveal the propagation of gravitational waves sent
forth by the planet in its motion around the Sun: indeed, this
motion does not generate any gravitational radiation; \emph{iii})
the binary RX J0806.3+1527 has the shortest known revolution
period (only 321 s): however, it cannot be a candidate for the
detection of gravitational waves because no kind of motion of a
mass can give origin to a gravitational wave.
\end{abstract}

\vskip1.20cm
%\section{Introduction}
\noindent {\bf Introduction}\par \vskip0.10cm
 Innumerable papers have been written on the Einsteinian orbits of the celestial
 bodies. For clear reasons, the overwhelming majority of them are
 of a perturbative character. In the present Note I consider the
 above orbits only so far as the hypothesized emission of
 gra\-vi\-ta\-tio\-nal waves is concerned. Now, we shall see that from
 this standpoint it is possible to develop some very simple and
 general considerations from which one can conclude that
 \emph{no motion of the celestial bodies gives origin to a gravitational
 radiation}.

\vskip0.80cm
%\section{}
\noindent {\bf 1.}-- {\bf Theory} \par \vskip0.10cm Several
arguments prove that no motion of masses can generate
gravitational waves \cite{1}. An essential demonstration may be
resumed in the following way. \par Consider a \emph{continuous
medium} characterized by whatever mass tensor $T_{jk}(x)$,
$(j,k=0,1,2,3)$, and let $g_{jk}(x)$ be the solution of the
Einstein field equations corresponding to a generic motion with
respect to a given reference system $x \equiv
(x^{0},x^{1},x^{2},x^{3})$. Suppose to follow ideally the motion
of a given \emph{mass element} describing a world line $L$, and
suppose that at a given time this element begins to emit a
gravitational wave.  Now, if we refer its motion to some
Riemann-Fermi coordinates $z \equiv (z^{0},z^{1},z^{2},z^{3})$
\cite{2}, the components $h_{jk}(z)$ of the metric tensor will be
equal to some \emph{constants} for \emph{all} points of $L$. This
means that the gravitational field \emph{on}$L$ can be
obliterated: consequently, no gravitational wave can be actually
sent forth by the considered mass element. But line $L$ is quite
generic, and therefore no motion of the medium gives origin to a
gravitational radiation.
\par It is implicit in the above reasoning that no gravitational
damping force has influenced the motion of our mass element.

\vskip0.80cm
%\section{}
\noindent {\bf 2.}-- {\bf On the binary PSR1913+16} \par
\vskip0.10cm According to many authors, an indirect experimental
proof of the phy\-si\-cal existence of the gravitational waves is
given by the time decrease of the orbital period $P_{b}$ of the
binary radiopulsar PSR1913+16 \cite{3}. The measured value of
$\textrm{d}P_{b}/\textrm{d}t$ is $-(2.30 \pm 0.22) \cdot
10^{-12}$, while the quadrupole formula of the linearized
relativity gives a $\textrm{d}P_{t}/\textrm{d}t$ due to a
hypothesized emission of gra\-vi\-ta\-tio\-nal radiation equal to
$-2.4 \cdot 10^{-12}$. An excellent agreement with the
observational data has been also obtained by computations at third
order of the gravitational constant and fifth order in $v/c$.
However, this agreement is rather suspect owing to the
unreliability of the adopted perturbative treatments
(cf.\cite{4}).
\par
As a matter of fact, we know from the theoretical proof of
sect.{\bf 1} that no motion of the stars of the above binary
system can generate gravitational waves. Accordingly, it would be
necessary to re-examine carefully the influence on
$\textrm{d}P_{b}/\textrm{d}t$ of those realistic effects that
Damour and Taylor have discarded as scarcely significant \cite{5}.
\par On the other hand, the computations of Taylor \emph{et alii}
are based on the assumption that the two stars of the system
PSR1913+16 can be treated as \emph{point} masses so far as their
motions are concerned: now, point masses interacting only
gravitationally describe \emph{geodesic} lines (cf. the first
paper cited in \cite{1}), i.e. their motions are "free",
"inertial" motions \emph{with} \emph{\textbf{no}} \emph{radiation
damping}.
\par The conclusion is obvious: the binary PSR1913+16 gives no
experimental proof of the real existence of the gravity waves.

\vskip0.80cm
%\section{}
\noindent {\bf 3.}-- {\bf On the motion of Jupiter}
\par \vskip0.10cm
As is well known, as a consequence of the quadrupole formula of
the linearized relativity, the power of the gravitational
radiation sent out by Jupiter in its motion around the Sun is
$\approx450$ watt. (The power of the solar \emph{electromagnetic}
radiation is about $10^{24}$ times greater). From the standpoint
of the \emph{exact} theory, this result is a pure nonsense because
-- cf. sect.{\bf 1} -- no motion of a body generates gravitational
waves. However, the present astrophysical community gives an
excessive credit to the linearized version of general relativity
and to perturbative computations starting from it. Thus, in recent
times we have seen a proclamation according to which an indirect
experimental detection of the gravitational waves emitted by
Jupiter in its motion would be quite possible, see Kopeikin and
Fomalont \cite{6}. These authors declare that, owing to a rare
alignment of Jupiter's motion against the quasar J0842+1835,
measurements of the propagation of the quasar radio-signals past
Jupiter must be sensitive to the propagation of the gravitational
radiation emitted by the planet. However, Asada \cite{7} and
subsequently Will \cite{8}, by means of more reasonable
perturbative treatments, have proved the non-existence of the
Kopeikin-Fomalont effect.
\par Of course, this result is obvious from the rigorous point of
view of the exact formulation of general relativity.

\vskip0.80cm
%\section{}
\noindent {\bf 4.}- {\bf On the binary RX J0806.3+1527}
\par \vskip0.10cm This binary system is composed of two white dwarfs
revolving around each other at a distance of only $8 \cdot 10^{4}$
km. The speed of the orbital motion is over $10^{3}$km/s, the
orbital period amounts to 321 s: it is the shortest known
revolution period, see Israel \emph{et alii} \cite{9}. These
authors believe that the above binary is an excellent candidate
for the detection of the gravitational waves, owing to the
shortness of its period. According to them, the space experiment
LISA (Laser Interferometer Space Antenna), that will be launched
within 2020, will be able to reveal the gravitational waves
emitted by RX J0806.3+1527. Of course, this is a pure wishful
thinking, see sect.{\bf 1}.
\par In conclusion, I desire also to emphasize that the
astrophysical community "ignores" that the so-called gravitational
waves are mere mathematical undulations devoid of a real energy
and a real momentum \cite{10}.

\small \vskip0.5cm\par\hfill {\emph{``}\emph{Sir, I have found you
an argument:}
  \par\hfill \emph{I am not obliged to find you}
  \par\hfill \emph{an understanding.''}
   \vskip0.10cm\par\hfill \emph{Samuel Johnson}}


\normalsize


\small
\begin{thebibliography}{99}

\bibitem{1}
A. Loinger, \emph{Nuovo Cimento} B, \textbf{115} (2000) 679; Idem,
physics/0106052, June 17th, 2001 (misclassified, proper
classification: gr-qc); Idem, \emph{Spacetime \& Substance},
\textbf{3} (2002) 129. See also A. Loinger, \emph{On Black Holes
and Gravitational Waves}, (La Goliardica Pavese, Pavia) 2002, Part
II.

\bibitem{2}
H. Weyl, \emph{Kommentar zu Riemanns ``Uber die Hypo\-the\-sen,
welche der Geometrie zugrunde liegen''}, 3. Auf\-lage (J.
Springer, Berlin) 1923, sect.3; T. Levi Civita, \emph{Mathem.
Annalen} \textbf{97} (1926) 291; also in \emph{Opere
ma\-te\-ma\-ti\-che -- Memorie e Note}, Vol.$4\,^{\circ}$
(Zanichelli, Bologna) 1960, p.433.

\bibitem{3}
R.A. Hulse, \emph{Revs. Modern Phys.}, \textbf{66} (1994) 699;
J.H. Taylor, \emph{ibidem}, \textbf{66} (1994) 711; and the
bibliography cited there. See also B.F. Schutz, \emph{Class.
Quantum Grav.}, \textbf{16} (1999) A131.

\bibitem{4}
A. Loinger, \emph{Spacetime \& Substance}, \textbf{3} (2002) 145.

\bibitem{5}
T. Damour and J.H. Taylor, \emph{Astrophys. J.}, \textbf{366}
(1991) 501.

\bibitem{6}
S.M. Kopeikin and E. Fomalont, gr-qc/0206022 v1, June 7th, 2002;
also in \emph{Proceedings of the 6th European VLBI Network
Symposium -- Ros E., Porcas R.W., Zensus J.A. (eds.) -- June
25th-28th 2002, Bonn, Germany}. See also K. Sawyer,
\emph{Washington Post}, Ja\-nua\-ry 8th, 2003, p.A03.

\bibitem{7}
H. Asada, astro-ph/0206266 v1, June 17th, 2002.

\bibitem{8}
C.M. Will, astro-ph/0301145 v1, January 9th, 2003.

\bibitem{9}
G.L. Israel \emph{et alii}, astro-ph/0203043 v1, March 4th, 2002;
to be published in \emph{Astronomy and Astrophysics Letters}; see
also ESO Press Release 06/02, March 15th, 2002.

\bibitem{10}
T. Levi-Civita, \emph{Rend. Acc. Lincei}, \textbf{26} (1917) 381;
also in \emph{Opere matematiche -- Me\-mo\-rie e Note},
Vol.$4\,^{\circ}$ (Zanichelli, Bologna) 1960, p.47. For an English
translation see physics/9906004, June 2nd, 1999.

\end{thebibliography}
\end{document}